\documentclass[11pt,preprint]{revtex4-1}

\usepackage[normalem]{ulem} 
\usepackage{amsmath,bm}
\usepackage{amsfonts}
\usepackage{amssymb,enumerate}
\usepackage{graphicx,hyperref}
\usepackage{amssymb}
\usepackage{epstopdf,booktabs}
\usepackage[usenames,dvipsnames]{color}
\usepackage[T1]{fontenc}
\usepackage[utf8]{inputenc}
\usepackage{natbib}
\newcommand{\be}{\begin {align}}
\newcommand{\ee}{\end {align}}
\newcommand{\beqa}{\begin {eqnarray}}
\newcommand{\eeqa}{\end {eqnarray}}

\hypersetup{
    colorlinks,    citecolor=blue,    filecolor=blue,    linkcolor=blue,    urlcolor=blue
}

\begin{document}
\title{Strong energy enhancement in a laser-driven plasma-based accelerator through stochastic friction}
\author{Z. Gong$^{1,2}$}
\author{F. Mackenroth$^{3,4}$}
\author{X. Q. Yan$^{1}$}
\author{A. V. Arefiev$^{4}$}
\email[corresponding author, ]{aarefiev@eng.ucsd.edu}
\affiliation{$^1$SKLNPT, School of Physics, Peking University, Beijing 100871, China}
\affiliation{$^2$Center for High Energy Density Science, The University of Texas at Austin, Austin, TX 78712, USA}
\affiliation{$^3$Max Planck Institute for the Physics of Complex Systems, 01187 Dresden, Germany}
\affiliation{$^4$University of California at San Diego, La Jolla, CA 92093, USA}


\date{\today}

\maketitle
%
{\textbf{Conventionally, friction is understood as an efficient dissipation mechanism depleting a physical system of energy as an unavoidable feature of any realistic device involving moving parts, e.g., in mechanical brakes. In this work, we demonstrate that this intuitive picture loses validity in nonlinear quantum electrodynamics, exemplified in a scenario where spatially random friction counter-intuitively results in a highly directional energy flow. This peculiar behavior is caused by radiation friction, i.e., the energy loss of an accelerated charge due to the emission of radiation. We demonstrate analytically and numerically how radiation friction can enhance the performance of a specific class of laser-driven particle accelerators. We find the unexpected directional energy boost to be due to the particles' energy being reduced through friction whence the driving laser can accelerate them more efficiently. In a quantitative case we find the energy of the laser-accelerated particles to be enhanced by orders of magnitude.}}

For an accelerated particle of charge $q$ and mass $m$ the main energy dissipation, or friction, is the continuous emission of electromagnetic radiation, referred to as \textit{radiation friction} (RF). The energy loss per unit time is given by~\cite{jackson1999classical}
\begin{align}
 P = \frac{2q^2\varepsilon^2}{3m^4c^7} \left(\frac{dp^\mu}{dt}\right)^2, \label{eq:Larmor}
\end{align}
where $c$ is the speed of light, $p^\mu$ the particle's relativistic momentum, $\varepsilon$ its energy and $t$ time. Several structural peculiarities of RF were described, e.g., mathematically ill-behaved particle dynamics~\cite{dirac1938classical} or the need for charge renormalization in classical electrodynamics~\cite{landau2013classical}. For RF to reduce a particle's energy $\varepsilon$ at a rate corresponding to the particles' momentum change due to acceleration, the emitted power $P=d\varepsilon/dt=(d\varepsilon/d|\bm{p}|)(d|\bm{p}|/dt)$, with the particle momentum $\bm{p}$, needs to match the accelerating force, resulting in $\left|\bm{F}_\text{RF}\right|=\left|d\bm{p}/dt\right|\sim 3m^4c^8/2\varepsilon^2q^2$. Early studies deemed this regime of \textit{instantaneous RF}, unreachable in a lab, as it would require accelerating electromagnetic fields $E_\text{RF} \geq F_\text{RF}/|e| \sim 10^{11}$ statV/cm even for the lightest particle, an electron (charge $-e<0$, mass $m_e$) at mildly relativistic energies of $\varepsilon_e \sim 100 m_e c^2$. Such field strengths, however, are becoming increasingly available due to the advent of ultra-high intensity lasers~\cite{strickland1985_CPA,mourou2006optics} providing intensities $I_L\gtrsim 10^{22}\, \text{W}/\text{cm}^2$ at optical wavelengths ($\lambda_L\sim1\,\mu$m), corresponding to electric fields $E_L \gtrsim 10^{11}$ statV/cm~\cite{Danson_etal_2015}, with facilities aiming at higher fields under construction \cite{ELI_WhiteBook}, sparking renewed interest in instantaneous RF~\cite{di2012extremely,bulanov_RDR}. At these facilities even experimental observation of instantaneous RF became possible \cite{RR_2018experimental_PRX,RR_poder2018experimental_PRX} in the random energy loss of a laser-accelerated relativistic electron bunch when scattered by a high-power laser pulse. Similar setups were investigated in a series of theoretical studies~\cite{di2009strong,thomas2012strong,harvey2012intensity,neitz2013stochasticity,green2014transverse,Anomalous_RT,vranic2014all,li2014robust,dinu2016quantum,harvey2017quantum}, indicating the acceleration of electrons as an important field of application for high-intensity laser facilities~\cite{pukhov2002strong,RMP_LWFA_esarey_2009,arefiev2016_beyond}. 
%
%
\begin{figure}[t]
\centering
\includegraphics[width=\linewidth]{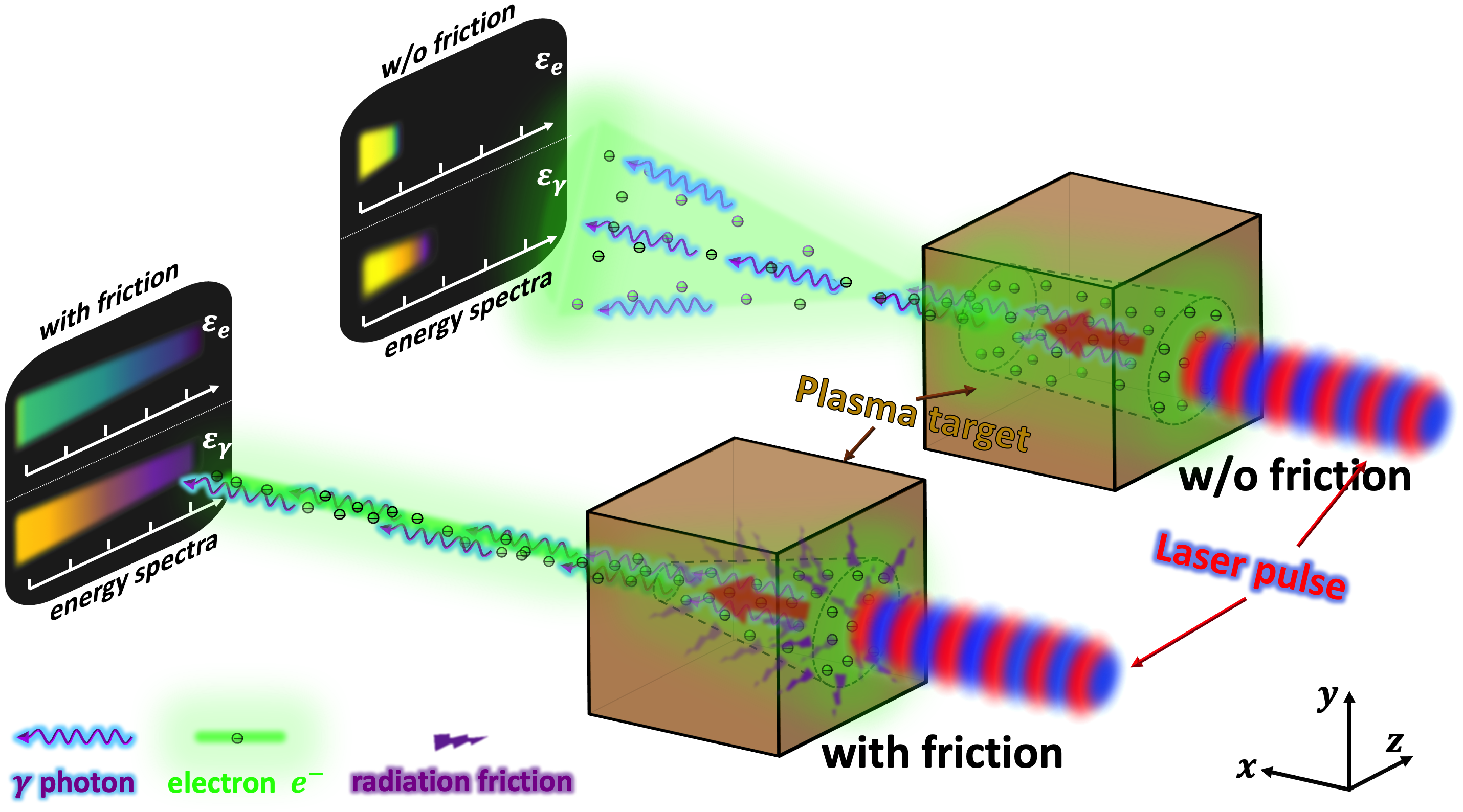}
\caption{Schematic diagram, the interaction between the ultra-intense laser pulse and a plasma channel target, picturing electrons (green dots), gamma-photons (blue wavy lines) and RF damping (purple fragments), without RF (upper/right panel) and with RF (lower/left panel) accounted for. Red arrows inside the plasma indicate the direction of the electron beam's acceleration and the black dashed lines indicate its outlines. Spectra of electron and photon energies, $\varepsilon_e$ and $\varepsilon_\gamma$, respectively, (black panels to the left) and their beam divergence (cones left of plasma targets) visualize the pronounced enhancement in peak energies and beam collimation with RF taken into account.}
\label{schematic}
\end{figure}
%
%
In contrast to the previous investigated acceleration mechanisms, that neglected RF, and in addition to the above mentioned peculiarities, in this work we uncover another counter-intuitive feature of instantaneous RF: We describe and characterize how RF can result in a significantly enhanced directional energy flow of laser-accelerated electrons, instead of energy dissipation, unveiling not only a novel peculiarity of RF but also providing an innovative optimization of laser-electron accelerators.

We study an ultra-strong laser pulse hitting a solid target, immediately ionizing it to a plasma with electron density $n_e$ (s.~Fig.~\ref{schematic}). We consider densities close to the critical threshold $n_c=\pi m_e c^2 /(\lambda_L^2 e^2)$, above which radiation cannot propagate inside the plasma. In an ultra-intense laser pulse, however, electrons oscillate relativistically, yielding an effective mass increase, increasing the critical density by a factor $\overline{\varepsilon}_e/m_ec^2=\sqrt{1+a_0^2}$, for an initially cold, laser-driven plasma \cite{Akhiezer_Polovin_1956,Predhiman_Dawson_1970,RT_2012_np,RT_2017_pop}, with the normalized laser amplitude $a_0\equiv \sqrt{I_L[\text{W/cm$^2$}]/1.37\times10^{18}}\lambda_L[\mu m]$. For $a_0\gg1$, i.e., $I_L\gg 10^{18} \text{W}/\text{cm}^2$ for $\lambda_L\sim 1\, \mu$m, the laser can penetrate the target and drive a volumetric electron current $J_0$ in an ensuing plasma channel. 
This current, in turn, generates a strong, quasi-static azimuthal magnetic field \cite{nakamura_MVA_2010_PRL,bulanov2010generation,Stark_PRL_2016,jansen2018_PPCF}, which was found to efficiently guide and collimate electron beams~\cite{pukhov1999_DLA,liu_2013_PRL_DLA,FSSA_2018}. 
An analytical test particle model [s. Methods, eq.~(\ref{eq:Methods_EnergyGain})] predicts that, when RF is accounted for, an electron inside the plasma channel can gain energy up to the maximum threshold
\begin{align}
  \varepsilon_e^\text{max} \approx 5.6 \left(\frac{a_0}{a_{MB}^3}\right)^{\frac14}\, \text{GeV}, \label{eq:MaxEnergy}
\end{align}
where $a_{MB}$ is the maximum potential of the channel magnetic field [s. Methods]. The energy gain is due to the electron staying in an accelerating laser phase over a long distance leading to an efficient acceleration. In contrast, without accounting for RF, quick dephasing between the electron and the laser pulse prevents comparable energy gain. Hence, RF is expected to significantly enhance the electrons' longitudinal momenta as well as the beam collimation.


To closely resemble conditions found in self-consistent plasma simulations~\cite{Stark_PRL_2016,FSSA_2018}, we study an electron with initial momentum $p_0 = 100 m_e c$. The driving laser has a normalized amplitude $a_0=200$ and wavelength $\lambda_L=1\, \mu$m, corresponding to an optical laser intensity of $I_L\approx5\times 10^{22}\, \text{W}/\text{cm}^2$, well within the reach of next-generation laser facilities~\cite{ELI_WhiteBook}, propagating along the $x$-axis and linearly polarized in $y$-direction. Furthermore, we assume a normalized current density $\alpha=0.03$, corresponding to an electric current $J_0\approx 3.2$ MA within a radius $r=4\mu m$. With these values our theory predicts substantial energy gain [s. Methods, eq.~(\ref{eq:LimitEnergyGain})]
\begin{align}
  \frac{d\varepsilon_{e}}{dt} \sim 10^{-5}\, a_0 \left[\frac{\text{GeV}}{\text{fs}}\right], \label{eq:energychange}
\end{align}
\begin{figure}[t]
\centering
\includegraphics[width=0.98\linewidth]{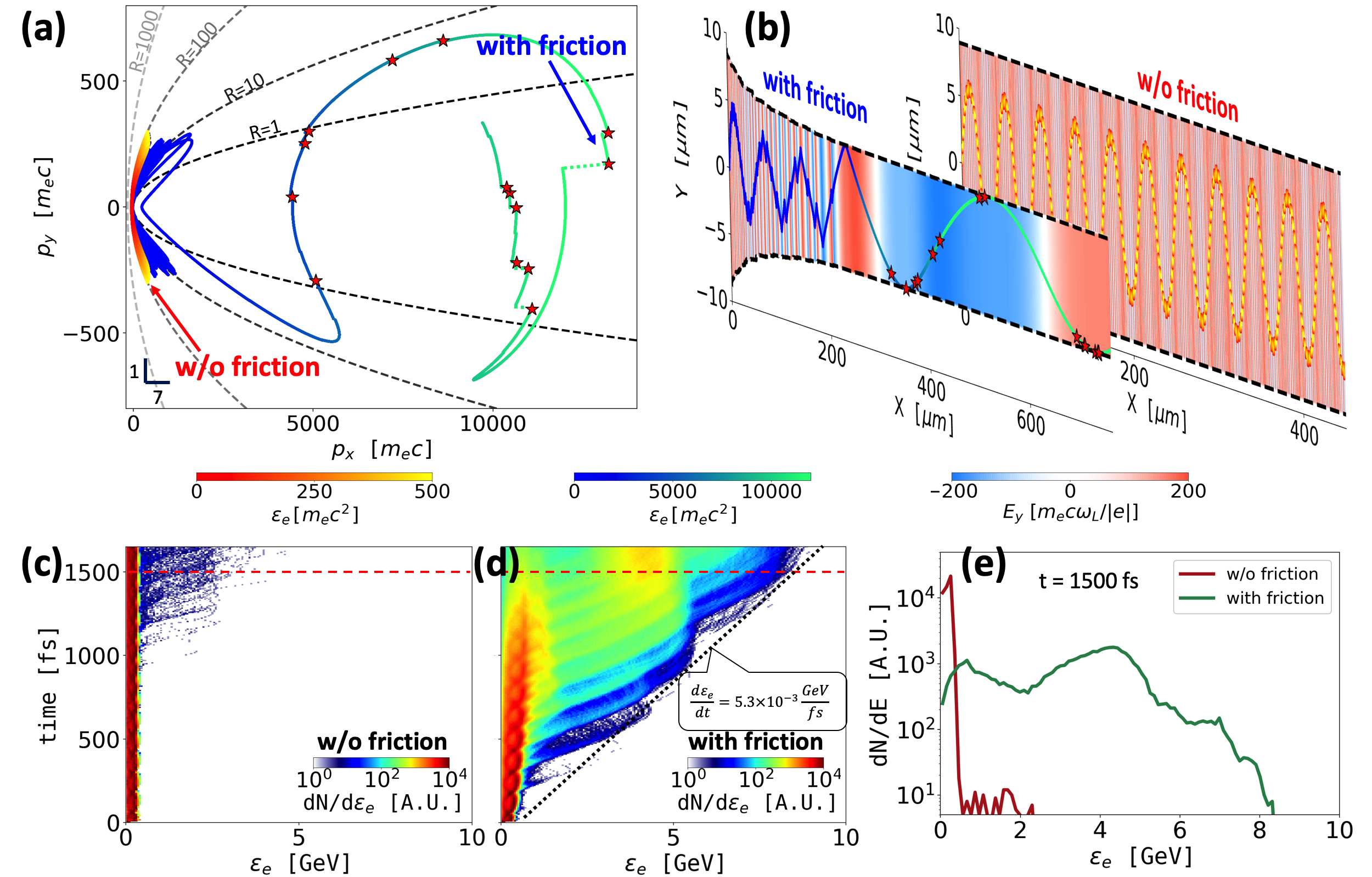}
\caption{(a): Electron trajectories in $p_x$-$p_y$ momentum-space without and with RF (instantaneous energy in red-yellow and blue-green colorbar, respectively). Dashed gray lines indicate analytical momentum space trajectories for different dephasing values $R$ and stars indicate photon emission with energy $\varepsilon_\gamma>50$ MeV. (b): Same electron trajectories as in (a) in $x$-$y$ coordinate-space (energy color-coded as in (a)), magnetic boundary $y_\text{MB}(t)$ (dashed black lines) and transverse electric laser field at the electron's position (red-blue colorbar). (c): Time resolved electron spectra without  RF included. (d): Same as (c) with RF included and a linear fit to the peak electron energy gain (dotted black line in). (e): Electron energy spectra at time t = 1500 fs (dashed red lines in (c),(d)).
}
\label{electron_comb}
\end{figure}
up to a maximal energy of $\varepsilon_e^\text{max} \sim 10 $ GeV, according to eq.~(\ref{eq:MaxEnergy}) where from numerical simulations we found it reasonable to assume $a_{MB}\sim10^{-2} a_0$, due to RF losses. To test these predictions, we study an ensemble of $10^6$ electrons with typical initial momenta perpendicular to the laser's propagation direction uniformly distributed in $p_0/m_ec \in \left[40, 160\right]$ in a numerical test particle model simulating the full single electron dynamics with RF accounted for as discrete quantum stochastic recoil [s. Methods]. When RF is neglected the electron oscillates at low longitudinal momentum (s.~Fig.~\ref{electron_comb}(a)). On the other hand, accounting for RF the electron gains significant longitudinal momentum. Investigating the two exemplary trajectories in position space (s.~Fig.~\ref{electron_comb}(b)) it becomes apparent that, when RF is accounted for, the electron gains energy when it approaches $y_\text{MB}$ and stays in the laser phase accelerating it along its propagation direction for a long time (blue area in left panel in Fig.~\ref{electron_comb}(b)). Without accounting for RF, on the other hand, dephasing prevents energy gain. Time-resolved spectra of the energies of the full electron ensemble further corroborate the beneficial effect of RF as a large number of electrons is accelerated to high energies with an approximately linear energy gain with time of $d\varepsilon_e/dt \approx 5.3\times 10^{-3} \text{GeV}/\text{fs}$ in close agreement with the analytical prediction of eq.~(\ref{eq:energychange}) (s.~Fig.~\ref{electron_comb}(c),(d)). We find this linear trend stable even after a long interaction time of $t=1500$ fs, resulting in a broad spectrum stretching to cutoff energies of $\varepsilon_e\approx8$ GeV (s.~Fig.~\ref{electron_comb}(e)), in reasonable agreement with the prediction of eq.~(\ref{eq:MaxEnergy}). In contrast, when neglecting RF no substantial part of the electron ensemble is accelerated to high energies (note log-scale in Fig.~\ref{electron_comb}(c)-(e)).
\begin{figure}[t]
\centering
\includegraphics[width=0.98\linewidth]{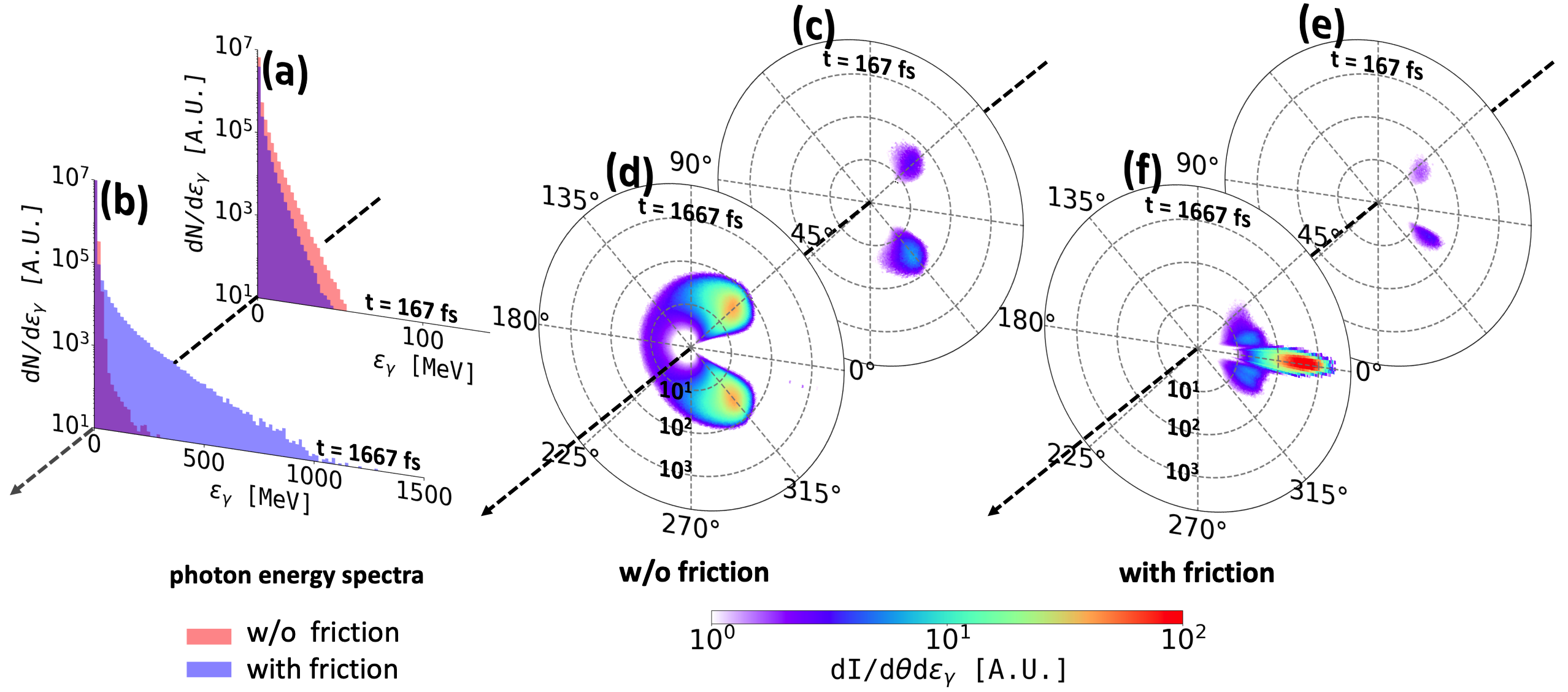}
\caption{(a)-(b): Energy spectra of the emitted photons with RF (purple) and without (red) at times 50 T$_L$ (a) and 500 T$_L$ (b) with $T_L=\lambda_L/c=3.33$fs. (c)-(d): Angular distribution of emitted photons without RF with the radial coordinate denoting the photon energy $\varepsilon_\gamma$ in log scale and the rotating coordinate the polar angle $\theta = \text{tan}^{-1}(p_{\gamma,z}/p_{\gamma,x})$ at the same time instants as (a)-(b). (e)-(f): Same as (c)-(d) but with RF taken into account.
}
\label{photon_comb}
\end{figure}
In addition, the electron's gamma-ray emission is well approximated by eq.~(\ref{eq:Larmor}) and angularly confined within an angle $\Delta \theta \sim m_ec^2/\varepsilon_e$. Consequently, increased electron energies should lead to an enhanced gamma-ray emission and narrower collimation. Time-integrated energy spectra of the emitted photons reveal that while at early times RF does not cause significant differences in the emitted photon spectra (s.~Fig.~\ref{photon_comb} (a)), at later times the photon spectrum with RF stretches to a cut-off energy $\varepsilon_\gamma^\text{RF} \sim1$ GeV, as compared to $\varepsilon_\gamma^\text{noRF}\sim100$ MeV if RF is neglected (s.~Fig.~\ref{photon_comb} (c)). Furthermore, we find the integrated radiation signal at early times to be centered in two lobes at comparable angles $\theta_\gamma \approx \pm 45^\circ$, irrespective of whether RF is taken into account or not (s.~Fig.~\ref{photon_comb} (d),(g)), whereas at late times RF leads to a significant collimation of the emitted photons as compared to the case with RF neglected, in which the angular photon distribution preserves the two-lobe structure (s.~Fig.~\ref{photon_comb} (f),(i)).

In summary, we have identified an experimentally realizable setup in which the conventional concept of friction is reversed: Instead of an undirected loss of energy, in the studied class of laser-electron accelerators realized in relativistically underdense plasma channels, the inclusion of RF yielded a highly directional energy boost. We found the electron beam as well as the emitted gamma-rays to feature significantly higher energies and collimation. This counterintuitive effect was explained as a facilitated coupling to the driving laser due to the electrons' undirected radiative losses resulting in prolonged motion in phase with the driving laser, facilitating highly directional energy uptake. Our numerical studies suggest that this effect could be observed at next-generation laser facilities.

\section*{Methods}

\textbf{Modeling: electron dynamics - } The considered effect of RF is directly linked to the electron dynamics in the combined laser and plasma fields present in the channel~\cite{Stark_PRL_2016,FSSA_2018}. To study these dynamics, we make use of a simplified model capturing all essential physics, as demonstrated in detailed kinetic plasma simulations~\cite{Stark_PRL_2016, arefiev2016_beyond, FSSA_2018,jansen2018_PPCF}. These simulations motivate the following fundamental assumptions underlying the model: (i) Since transverse electron oscillations are confined narrower than the width of the laser beam the electron effectively experiences the laser as a plane wave. (ii) Since laser beam diffraction is suppressed in structured targets~\cite{Stark_PRL_2016,FSSA_2018}, the laser maintains its peak amplitude over distances much longer than its Rayleigh range. (iii) Dephasing between the electrons and the laser is primarily determined by the longitudinal electron velocity $v_x$ whence for the laser's wavelength and frequency $\omega_L$ we neglect deviations of the phase velocity $v_{ph} = \lambda_L\omega_L/2\pi$ from $c$, possibly arising due to the laser's propagation through the plasma. (iv) The laser drives a strong, uniform longitudinal electron current density $j_0 < 0$, with the return current flowing radially outside of the plasma channel. (v) Binary collisions, electric charge separation fields and feedback of the accelerated electrons on the bulk plasma dynamics are negligible. Consequently, the electron dynamics can be modeled by considering a single electron in a combined plane electromagnetic wave and a static azimuthal magnetic field representing the laser and magnetic plasma field, respectively.

We model the laser pulse to be linearly polarized along the $y$-axis and to propagate along the $x$-axis with electric and magnetic field $\displaystyle \bm{E}_{wave} = - m_e c/|e| \partial \bm{a}_{wave}(\xi)/\partial t$ and $\displaystyle \bm{B}_{wave} = m_e c^2/|e| \nabla \times \bm{a}_{wave}(\xi)$, respectively. Here $\xi \equiv (ct - x)/\lambda_L$ and the normalized vector potential $\bm{a}_{wave}$ has only one component, $\bm{a}_{wave} (\xi)= (0,a_0,0)\sin(2\pi\xi)$. The quasi-static magnetic plasma field sustained by the constant current density $j_0$, on the other hand, can be written as $\displaystyle \bm{B}_\text{channel} = m_e c^2/|e| \nabla \times \bm{a}_\text{channel}$ with $\bm{a}_\text{channel} = \bm{e}_x \alpha (y^2 + z^2)/\lambda_L^2$, where $\bm{e}_x$ is a unit vector. Here we introduced a dimensionless current density $\displaystyle \alpha \equiv -\lambda_L^2 j_0 / (4\pi J_A)$ with the Alfv\'en current $J_A \equiv m_e c^3 / |e|$. Since the laser pulse drives flat electron trajectories in the laser's polarization plane~\cite{FSSA_2018}, from now we consider $z\equiv0$ and study electron motion in the $(x,y)$-plane only. In the combined fields, $\bm{E} = \bm{E}_{wave}$ and $\bm{B} = \bm{B}_{wave} + \bm{B}_\text{channel}$, the electron dynamics with radiation friction (RF) taken into account are governed by the equations
\begin{eqnarray}
&& \frac{d {\bm{p}}}{d t} = - |e| {\bm{E}} - \frac{c|e|}{\varepsilon_e} \left[ {\bm{p}} \times {\bm{B}} \right] - F_\text{RF} \frac{\bm{p}}{\left|\bm{p}\right|}, \label{EQ_1} \\
&& \frac{d {\bm{r}}}{d t} = \frac{c^2\bm{p}}{\varepsilon_e} , \label{EQ_2}
\end{eqnarray}
where ${\bm{r}}$ and ${\bm{p}}$ are the electron position and momentum, respectively, $t$ is the time and $F_\text{RF} = \kappa \varepsilon_e^2 \mathcal{E}^2/m_e^2c^4$ quantifies the impact of RF where $\kappa=8\pi e^2/3\lambda_Lm_ec^2$. Typically, the electrons enter the plasma channel at time $t_0$ on-axis with a large transverse momentum~\cite{Stark_PRL_2016,FSSA_2018}, i.e., $y(t_0)=0$, $\bm{p}(t_0)=(0,p_{y,0}\gg m_ec,0)$. Then, since we study relativistic motion we retain only terms of leading order in $\varepsilon_e \gg m_ec^2$ in the dynamical RF parameter which can then be expressed as~\cite{landau2013classical}
\begin{align}
 \mathcal{E}^2 = \left(\frac{e}{m_ec\omega_L}\right)^2 \left(\bm{E}_\perp + \frac{c}{\varepsilon_e} \left[\bm{p}\times \bm{B}\right]\right)^2, \label{eq:DynamicParameter}
\end{align}
where $\bm{E}_\perp = \bm{E} - \bm{p}\left(\bm{p}\bm{E}\right)/\bm{p}^2$ is the electric field component perpendicular to the electron momentum. Combining the $x$ and $y$ components of eq.~(\ref{EQ_1}) and taking into account that $\bm{a}_{wave}$ is a function of $\xi$ alone and $\bm{p}\approx \varepsilon_e/c$ we find the following relation
\begin{align}
\frac{d \left(R + a_\text{channel}\right)}{d t} &= - F_\text{RF} \frac{d\xi}{dt},\label{eq:ConstantOfMotion}
\end{align}
where we introduced the \textit{dephasing rate} $R=(d\xi/dt)\varepsilon_e \lambda_L/m_ec^3=(\varepsilon_e-cp_x)/m_ec^2$ quantifying the change of $\xi$ at the electron's instantaneous position. Hence, for vanishing RF ($F_\text{RF}\to0$) the quantity $R+a_\text{channel} = \text{const}. + \mathcal{O}\left(F_\text{RF}\right)$ is an integral of motion, which, in turn, implies that the motion of a relativistic electron with the above introduced initial conditions is confined to transverse displacements smaller than the \textit{magnetic boundary} $y\leq y_{MB}:=\lambda_L\sqrt{p_{y,0}/m_e c \alpha}$. Taking into account RF, however, one finds that for the same initial conditions the magnetic boundary shrinks as a function of time according to
\begin{align}
 y_{MB}(t)=\lambda_L\sqrt{\frac{p_{y,0}}{m_ec\alpha}- \int_{\xi(t_0)}^{\xi(t)} d\xi\ \frac{F_\text{RF}}{\alpha}}. \label{eq:MagBoundary}
\end{align}
Since the channel potential $\bm{a}_\text{channel}$ increases quadratically in $y$ this, in turn, implies that it is limited by $\left|\bm{a}_\text{channel}\right|\leq a_{MB} = \left|\bm{a}_\text{channel}\left[y_{MB}(t)\right]\right|$. 

\textbf{Modeling: energy gain - } We now turn to studying the electron's energy gain. From eq.~(\ref{EQ_1}) we see that the electron's energy gain is given by the balance between acceleration and deceleration in the laser field and losses to radiation friction
\begin{align}
 \frac{d\varepsilon_e}{dt} = \sum_{i=x,y,z} \frac{p_ic^2}{\varepsilon_e} \frac{dp_i}{dt} = \frac{m_e c^3\omega_L}{\varepsilon_e}\bigg(p_y \cos\left(2\pi\xi\right)a_0 - \left|\bm{p}\right|F_\text{RF} \bigg). \label{eq:EnergyGain}
\end{align}
For RF being completely negligible, the electron's rate of energy gain in an optical laser field ($\lambda_L = 1\, \mu$m) is hence limited by 
\begin{align}
\frac{d\varepsilon_e}{dt} \lesssim \omega_L a_0 m_ec^2 \sim 10^{-5}\, a_0 \left[\frac{\text{GeV}}{\text{fs}}\right]. \label{eq:LimitEnergyGain}
\end{align}
On the other hand, we immediately see that for quick dephasing $R\gg1$, as is typical in the studied setup, the energy gain is a strictly periodic function, whence the electron cannot gain significant energy. Since, as apparent from eq.~(\ref{eq:ConstantOfMotion}), RF affects the dephasing, this dephasing symmetry can be broken by RF and the electron stay in an accelerating laser phase for a long time, leading to significant energy gain. To now quantitatively estimate the maximum energy the electron can gain we need an explicit expression for $\mathcal{E}$, which requires detailed knowledge of the electron motion in the $(x,y)$-plane. To this end, we note that once the electron is accelerated in the laser's propagation direction its velocity will only make a small angle $\theta\ll1$ with the $x$-axis and we write $\bm{p}=\left|\bm{p}\right|(\cos(\theta),\sin(\theta),0)$. For highly relativistic motion we can then rewrite $p_y = \left|\bm{p}\right|\sin(\theta) \lesssim \varepsilon_e\theta/c$ in eq.~(\ref{eq:EnergyGain}) to find the energy gain to be limited by $\displaystyle d\varepsilon_e/dt \lesssim a_0 m_ec^2 \omega \theta - \kappa \varepsilon_e^2 \mathcal{E}^2/m_ec^2$. We now distinguish the following two cases: (i) If the electron radiates energy mainly due to the laser field's action, then after some algebra one finds from eq.~(\ref{eq:DynamicParameter}) $\mathcal{E}^2 \approx \mathcal{E}^2_\text{laser}:=\left(\theta^4+\gamma^{-4}\right)a_0^2/4$. (ii) If, on the other hand, the radiative losses are dominated by the channel magnetic field one finds from eq.~(\ref{eq:MagBoundary}) $\mathcal{E}^2\approx \mathcal{E}^2_\text{channel}:=\left(e\left|\bm{B}_\text{channel}\right|/m_ec\omega\right)^2 \leq a_{MB}^2(t)$. Comparing these two cases we find that for electrons propagating at large angles with respect to the laser's propagation direction $\theta \geq \theta_B:=\sqrt{2a_{MB}/a_0}$ radiative losses are always dominated by the laser field. On the other hand, we find that even for small angles for $a_{MB}\leq a_{MB}^\text{min}=m_e^2c^4a_0/2\varepsilon_e^2$ the radiative losses are always dominantly due to the laser field. Since, for a typical situation $\varepsilon_e \sim m_ec^2 a_0\gg m_ec^2$ it is $a_{MB}^\text{min}\sim a_0^{-1}\ll1$, whereas in the studied setup it typically holds $a_{MB} > 1$. We are thus going to assume $a_{MB}\geq a_{MB}^\text{min}$ from now on, which implies $\theta_B\geq m_ec^2/\varepsilon_e$.

Now, for propagation at large angles $\theta\geq\theta_B\geq \gamma^{-1}$, i.e., the radiative losses being dominated by the laser field, the electron's energy gain is given by $\displaystyle \left.d\varepsilon_e/dt\right|_\text{laser} \approx 2\pi a_0 m_ec^2\theta \left[1 - \kappa \varepsilon_e^2 \theta^3 a_0/8\pi m_e^2c^4\right]$. Thus, the electron can gain energy only provided
\begin{align}
 \theta\leq\theta_\text{laser} := \left(\frac{8\pi m_e^2c^4}{\kappa a_0\varepsilon_e^2}\right)^{\frac13}. \label{eq:LaserAngle}
\end{align}
On the other hand, for propagation at small angles $\theta\leq \theta_B$, i.e., the radiative losses being dominated by channel magnetic field, the electron's energy gain is given by $\displaystyle \left.d\varepsilon_e/dt\right|_\text{laser} \approx 2\pi a_0 \theta - \kappa \gamma^2 a_{MB}^2$. Consequently, the electron can gain energy as long as
\begin{align}
 \theta\geq \theta_\text{channel} := \frac{\kappa a_{MB}^2\varepsilon_e^2}{2\pi m_e^2c^4a_0}. \label{eq:ChannelAngle}
\end{align}
We note a fundamental difference between eqs. (\ref{eq:LaserAngle},\ref{eq:ChannelAngle}): For propagation at large angles $\theta\geq\theta_B$ eq.~(\ref{eq:LaserAngle}) implies that the electron can gain energy only when reducing its propagation angle. For propagation at small angles $\theta\leq\theta_B$, however, eq.~(\ref{eq:ChannelAngle}) implies that the electron can gain energy only when increasing its propagation angle. This fundamental difference in scaling behavior, however, indicates that independent of its initial value the electron's propagation angle always tends to $\theta_B$. This asymptotic behavior, in turn, implies the existence of an upper boundary for the electron's energy beyond which energy gain is prohibited due to radiative losses
\begin{align}
 \varepsilon_e^\text{max}&= m_e c^2 \left(\frac{8\pi^2a_0}{\kappa^2 a_{MB}^3}\right)^{\frac14}, \label{eq:Methods_EnergyGain}
\end{align}
which can be equally obtained by either equating eqs. (\ref{eq:LaserAngle},\ref{eq:ChannelAngle}) or by equating either relation to $\theta_B$. Furthermore, for $\theta=\theta_B$ the energy gain in the laser- and channel-dominated regimes agree and from eq.~(\ref{eq:EnergyGain}) we see that for $\varepsilon_e \to \varepsilon_e^\text{max}$ it holds $\displaystyle d\varepsilon_e/dt \to 0$.

\textbf{Numerical particle dynamics - } The numerical model adopts a fourth-order Runge-Kutta algorithm to push the particle motion under the Lorentz force with the time step $\Delta t=5\times 10^{-4}\, T_L$ to satisfy the stringent temporal criteria not only on electron acceleration but also on photon generation. We model an infinite plane wave laser field by $E_{\text{wave},y}=a_0\cos(\xi)$ and $B_{\text{wave},z}=a_0\cos(\xi)$. The self-generated plasma magnetic field is given by $B_{\text{channel},z}=-2\alpha y$. Since gamma-ray emission to occur over a short distance, the emission probability is calculated under the local constant field approximation by the differential emission rate of an electron with energy $\varepsilon_e$ and quantum parameter $\chi_e=E_{rf}/E_\text{cr}$, where $E_{rf}$ is the electric field in the electron's instantaneous frame and $E_\text{cr}=m_e^2c^3/(e\hbar)\approx 1.3\times 10^{18}$V/m~\cite{Ritus_1985,bell2008possibility,duclous2010monte,ridgers2014modelling,gonoskov2015extended}
\begin{align}
    \frac{d^2N}{d\chi_\gamma dt}=\sqrt{3}\frac{m_ec^2}{h}\alpha_f\frac{\chi m_ec^2}{\varepsilon_e} \frac{F(\chi,\chi_\gamma)}{\chi_\gamma},
\end{align}
where we used the emitted photon's quantum parameter $\chi_\gamma=e\hbar|F_{\mu\nu} k^{\nu}|/m_e^3c^3$ and the fine structure constant $\alpha_f=e^2/4\pi\epsilon_0\hbar c\approx 1/137$. The radiated energy is given by~\cite{Ritus_1985}
\begin{align}
    F(\chi,\chi_\gamma) = \frac{4\chi_\gamma^2}{\chi^2}s K_{2/3}(s) + \left(1-\frac{2\chi_\gamma}{\chi}\right)s \int_s^\infty K_{5/3}(t)dt,
\end{align}
where $s=4\chi_\gamma/[3\chi(\chi-2\chi_\gamma)]$ and $K_{n}(s)$ are modified second order Bessel functions. 
Then each electron is initially assigned a final optical depth $\tau_f=\log[1/(1-P)]$, with a random number $P \in [0,1]$ modeling stochastic emission and straggling. The differential rate  
\begin{align}
 \frac{d\tau_\gamma}{dt}=\int_0^{\chi/2} \frac{d^2N}{dtd\chi_\gamma} d\chi_\gamma
\end{align}
is then advanced over each time step until the assigned optical depth is reached $\tau_\gamma \geq \tau_f$. In the corresponding time step, the electron emits a photon with its specific value $\chi_\gamma^f$ found from the relation 
\begin{align}
    \eta = \frac{\int_0^{\chi_\gamma^f} F(\chi,\chi_\gamma)/\chi_\gamma d\chi_\gamma}{\int_0^{\chi/2} F(\chi,\chi_\gamma)/\chi_\gamma d\chi_\gamma},
\end{align}
where $\eta\in [0,1]$ is a uniformly distributed random number. Then the photon energy, $\hbar\omega_\gamma$, is determined by $\hbar\omega_\gamma=2m_ec^2\chi_\gamma^f\gamma/\chi$ and the electron's momentum after the emission is given by $\vec{p}^{f}=[1-\hbar\omega_\gamma/(cp)]\vec{p}$. Finally, since the gamma-rays are primarily emitted within a cone of opening angle $\Delta \theta \lesssim m_ec^2/\varepsilon_e$ and we consider $\varepsilon_e\gg m_ec^2$, we assume them to be emitted along the electron's instantaneous direction of motion.

\section*{Data availability} All relevant numerical data supporting our findings are available from the corresponding author upon reasonable request.

%

\section*{Acknowledgments} 

The work has been supported by the National Science Foundation (Grant No. 1632777), the National Basic Research Program of China (Grant No.2013CBA01502), and NSFC (Grant No. 11535001).

\section*{Author contributions}

All authors together came up with the original idea, designed the setup and discussed the effect. Z.G. performed the numerical simulations, F.M. developed the analytical model, Z.G. and F.M. wrote the manuscript. A.V.A. guided and supervised the project.

\section*{Competing interests}

The authors declare no competing interests.

\section*{Materials \& Correspondence}

Correspondence and materials requests should be addressed to A.V.A.

\end{document}